\documentclass[
 reprint,
amsmath,amssymb,
aps,
showpacs,notitlepage,prb,aps,longbibliography]{revtex4-2}
\usepackage{graphicx}
\usepackage{bbm}
\usepackage{color}

\begin{document}

\title{The  closed piecewise uniform   string revisited}

\author{M. Bordag}
\email{bordag@uni-leipzig.de}
\affiliation{ Institute for Theoretical Physics, Universit{\"a}t Leipzig}

\author{I.G. Pirozhenko}%
\email{pirozhen@theor.jinr.ru}
\affiliation{BLTP, JINR Dubna and Dubna State University}

\newcommand{\eq}[2]{\begin{align}\label{#1}#2\end{align}}
\newcommand{\eqo}[1]{\begin{align*}#1\end{align*}}
\renewcommand{\Ref}[1]{(\ref{#1})}
\newcommand{\nn}{\nonumber}
\newcommand{\pa}{\partial}
\newcommand{\cb}{\color{blue}}
\newcommand{\Res}{\mathop{\rm res}\limits}
\newcommand{\sig}{\sigma}\newcommand{\ep}{\epsilon}
\newcommand{\al}{\alpha}\newcommand{\ga}{\gamma}
\newcommand{\om}{\omega}

\begin{abstract}
	We reconsider the composite string model introduced {30 years ago} to study the vacuum  energy. The model consists of a scalar field, describing the transversal vibrations of a string consisting of piecewise constant sections with different tensions and mass densities, keeping the speed of light constant across the junctions. We consider the spectrum using transfer matrices and Chebyshev polynomials to get a closed formula for the eigenfrequencies.  We calculate vacuum and free energy as well as the entropy of this system in two approaches, one using contour integration and another one using a Hurwitz zeta function. The latter results in a representation in terms of  finite sums over polynomials. Several limiting cases are considered as well, for instance, the high-temperature expansion, which is expressed in terms of the heat kernel coefficients. The vacuum energy has no ultraviolet divergences, and the corresponding heat kernel coefficient $a_1$ is zero due to the constancy of the speed of light. This is in parallel to a similar situation in macroscopic electrodynamics with isorefractive boundary conditions.
\end{abstract}

\maketitle


\section{\label{T1}Introduction}

In the piecewise uniform   string model~\cite{brev90-41-1185}, a closed or open string is considered, which consists of two or more segments.
This is a generalization of a homogeneous relativistic string with the same string tension $T$	everywhere.
In opposite to the standard case of a homogeneous string, this string consists of
homogeneous segments with different string tensions $T_i$ and mass densities, $\rho_i$, such that their ratio is just the constant speed of light,
\eq{1.2}{ c=\sqrt{\frac{T_i}{\rho_i}}.
}
We assume that alternating sections with different $T_i$ and $\rho_i$ repeat periodically and that the string is closed.

The canonical quantization of this model, which is also called {\it composite string model}, requires a  target space with the dimension $D=26$.  \cite{brev99-453-217}. However, in the string potential, the dimension  appears just as a factor $(D-2)$ in front of the  first quantum correction (the Casimir energy). The latter is calculated in (1+1)-dimensional space of the string world surface and coincides with the renormalized vacuum energy of a scalar field  which describes
the transversal vibrations of the string. This scalar field  $\phi(\sig,\tau)$ obeys the (1+1)-dimensional
 wave equation
\eq{1.1}{ \left(\frac{\partial^2}{\partial\tau^2}-
	\frac{\partial^2}{\partial\sigma^2}\right)\phi
	= 0,
}
with the matching conditions
\eq{1.1a}{ \phi|_{x=L-0} =\phi|_{x=L+0}, \qquad
\left.T_I	\frac{\partial \phi}{\partial\sigma}\right|_{x=L-0}
=\left.T_{II}\frac{\partial \phi}{\partial\sigma}\right|_{x=L+0},
}
at the junctions, which imply  the continuity of the displacement of the string and of the restoring force. Obviously, because of equations \Ref{1.2} and \Ref{1.1}, the model is relativistic.

In this model, we are primarily interested in the dispersion relation and the band structure of the string excitation spectrum.
Second, we consider the vacuum energy of excitations, that is, the Casimir effect associated with this system. Finally, we introduce  finite temperature and examine the corresponding thermodynamic quantities. The inclusion of impurities may be of additional interest. The merit of this model, which it shares with several others, is its simplicity, which makes it possible to study the physical quantities mentioned   most explicitly and easily.

The piecewise uniform string  was first considered in \cite{brev90-41-1185}, and shortly thereafter  in \cite{lixi91-44-560} a much simpler representation was found for the model, which in \cite{brev96-53-3224}  was generalized to a string with $2N$ equal sections.
In \cite{brev96-53-3224}, a string with three pieces was studied. Its vacuum energy has shown a non monotone (in opposite to the two-piece case, see below) dependence on the two  ratios of the tensions. We mention also an  open composite string model~\cite{hada00-62-025011}.

In the papers \cite{brev99-453-217} and ~\cite{brev03-44-1044}  the thermodynamic quantities were calculated  for a two-piece  and 2N-piece strings, and    the  Hagedorn (critical) temperature was found, which increases with the number of string segments and inverse proportional to $\sqrt{D-1}$.
The negative Casimir energy of the composite string has prompted some speculations about the importance of the model for cosmology~\cite{brev03-44-1044}.
In  \cite{bayi96-37-3662}, a twisted string was considered together with possible relations to processes in the early universe and to gravity.
A generalization to a charged string placed in  a magnetic field was studied in \cite{bern97-257-84}. In this paper also a variational principle for the string was set-up.
In \cite{brev99-40-1127}  an interesting  scaling property was found. It appears that the ratio of the vacuum energies, $f(x) = E_N(x)/E_N (0)$,  is approximately independent of $N$, provided $N\ge2$, and lies in the interval $0 < f(x) < 1$.

There are interesting links to the composite string model and  neighboring topics. Let us start with an analogy in macroscopic electrodynamics. If one considers some  material body having permittivity $\ep$ and permeability $\mu$, the speed of light inside is $c=1/\sqrt{\ep\mu}$. In such systems, the vacuum energy of the electromagnetic field has specific ultraviolet divergences which are even today not fully understood. For the first time this problem was observed in  \cite{schw78-115-1}, in detail it was investigated in \cite{bord99-59-085011}. This divergence is absent if the speeds of light inside and outside  the material bodies are equal. It must be mentioned that even for equal speeds of light across an interface, the electromagnetic fields are different, obeying  well-known matching conditions on the the interface. Boiling down to (1+1) dimensions, in the composite string model we observe  the same situation.

Further similar (1+1)-dimensional periodic objects are nanowires and optical lattices. There is a huge amount of papers on this topic and we let us mention only the first, \cite{kron31-130-499},  which introduced the Kronig-Penney model. Of special interest for their simplicity are also models with background potentials with zero range support; for a recent review see \cite{bord15-91-085038} and for the most general case in the sense of selfadjoint extensions see \cite{asor06-39-6127}.   There is  an  analogy between  closed composite $2N$ string discussed in the present paper and a  $\delta$-ring, which is a chain of $\delta$-potentials  placed on a  circle. In the recent papers ~\cite{bord1812.09022, bord20-35-2040005}, a formalism was put forward for calculating the vacuum energy  in one-dimensional periodic backgrounds formed of generalized $\delta$-potentials. The use of Chebyshev polynomials allowed to significantly simplify the calculations. In Sect. \ref{T2} we apply this approach to the 2N piecewise string.
In~\cite{shaj16-94-065003} Casimir  energies for  self-similar  (fractal-like) sets of parallel plates with $\delta$-potential separated by distances $z = a, \frac{a}{2},\frac{a}{4}, \frac{a}{8}...$ were studied. It was observed   that  these systems  manifest an analogy to the theory of the piecewise uniform string.

In most of the above mentioned papers on the composite string model, much effort was put into the investigations of  different regularizations of the vacuum energy of the string. However, looking from the point of the heat kernel expansion on these, one observes that the decisive  coefficient $a_1$ (which comes in (1+1)-dimensions in place of $a_2$ in (3+1))  is zero (see below in Sect. \ref{T43}). In such case, as known from the general theory (see, e.g., Chapt. 4 in \cite{BKMM}), all divergences can be removed uniquely, not leaving room for any ambiguity. In some regularizations, for instance, in the zeta functional one, there are no divergences at all. For this reason, we will be very brief on this topic.

Another intriguing point is negative entropy in Casimir effect like configurations, which  was first observed in \cite{geye05-72-022111}, for a recent overview see the introduction in \cite{milt17-96-085007} and for single standing objects see \cite{bord18-51-455001}, \cite{bord1807.10354} and \cite{bord20-80-221}.   Thus, it is necessary to consider the question of what entropy a piecewise  uniform string will show.

In the present paper, we reconsider the closed composite string and
calculate vacuum and free energies as well as the entropy using integral representation and sum representation as well.

We use zeta functional regularization and calculate the heat kernel coefficients, pointing out the uniqueness of the result. Also,  we get the high and low-temperature asymptotics of the free energy and demonstrating how the general scheme is applied in the given case.
We add not much, but we are interested in a more streamlined and most transparent and unified formulation of this topic.
\\
Throughout the paper we use units with $c=\hbar=k_{\rm B}=1$.

\section{\label{T2}The piecewise uniform  string and its spectrum}
We consider a  piecewise uniform closed string composed of $2N$ segments. After Fourier transform in $\tau$, the solution of the string equation,
\begin{equation}
(-\omega^2-\partial_\sigma^2)\phi(\sigma)=0,
\label{p1_1}
\end{equation}
consists of plane wave segments,
\begin{eqnarray}
\phi(\sigma)&=&\sum_{n}\phi_n(\sigma)\theta_n(\sigma),~~~
\nn\\\phi_n(\sigma) &=&A_n e^{i\omega (\sigma-a n)}+B_n e^{-i\omega(\sigma-a n)}
\label{p1_2},
\end{eqnarray}
and we use the notation
\begin{equation}
\theta_n(\sigma)=
\left\{
\begin{array}{cc}
1, & a (n-1)<\sigma <a n,\\
0, & \mbox{elsewhere}.
\end{array}
\right.
\label{p1_3}
\end{equation}
The limits of  fields and their derivatives  at the string junctions are denoted by
\begin{equation}
\phi_n^{\pm}=
\left.\left(
\begin{array}{c}
\phi_n(\sigma)\\
\phi_n'(\sigma)
\end{array}
\right)\right|_{a n\pm0}.
\label{p1_4}
\end{equation}
With the shorthand notation,
\begin{equation}
\Psi_n=\left(
\begin{array}{c}
A_n\\
B_n
\end{array}
\right),
\label{p1_6}
\end{equation}
the expression (\ref{p1_4}) can be rewritten as
\begin{eqnarray}
\phi_n^-&=&\left(
\begin{array}{c}
A_n+B_n\\
i\omega A_n-i\omega B_n
\end{array}
\right)=K \Psi_n ,
\label{p1_7}\\
\phi_n^+&=&\left(
\begin{array}{c}
A_n e^{-i\omega a}+B_n  e^{i\omega a}\\
i\omega A_n e^{-i\omega a} -i\omega B_n  e^{i\omega a}
\end{array}
\right)=K \, Q \, \Psi_{n+1} ,
\nn\label{p1_8}
\end{eqnarray}
with the matrices
\begin{equation}
K=\left(
\begin{array}{cc}
1 & 1\\
i\omega & -i\omega
\end{array}
\right), \quad
Q=\left(
\begin{array}{cc}
 e^{-i\omega a} & 0\\
0 &  e^{i\omega a}
\end{array}
\right).
\label{p1_9}
\end{equation}
The matching conditions \Ref{1.1a} at a junction can also be written using a matrix, $M_n$,
\begin{equation}
M_n \phi_n^-=\phi_n^+.
\label{p2_1}
\end{equation}
For the piecewise uniform string this matrix reads
 \begin{equation}
 M_n=\left(
 \begin{array}{cc}
 1 & 0\\
 0& x^{(-1)^{n+1}}
 \end{array}
 \right),
 \label{p2_3}
 \end{equation}
where $x$ is the ratio of the string tensions
\eq{2_3}{ x&=\frac{T_{I}}{T_{II}}.
}
For comparison, we mention  the matrix  corresponding  to  the most general delta-potential at the junctions
\eq{p2_2}{	M=\left(
	\begin{array}{cc}
		\frac{1-\beta}{1+\beta} & 0\\
		\frac{\alpha}{1-\beta^2}&\frac{1+\beta}{1-\beta}.
	\end{array}
	\right),
}
where $\al$ and $\beta$ are some couplings, as used, for example, in \cite{bord20-80-221}.

The transfer matrix $T_n$ is defined as to relate the solutions in neighboring segments,
\begin{equation}
T_n \Psi_n=\Psi_{n+1}.
\label{p2_6}
\end{equation}
Inserting (\ref{p1_7}) and (\ref{p1_8})  into (\ref{p2_1}),
\begin{equation}
M_n  K \Psi_n=K Q \,\Psi_{n+1},
\label{p2_7}
\end{equation}
we get
\begin{equation}
T_n= Q^{-1} K^{-1} M_n K.
\label{p2_8}
\end{equation}
 For two neighboring segments of the string, having with \Ref{p2_3}
\begin{equation}
 	\mbox{odd} \;  n:  ~ M_n=\left(
 	\begin{array}{cc}
 		1 & 0\\
 		0& x,
 	\end{array}
 	\right),
 	\quad
 	\mbox{even} \;  n:
 	~ M_n=\left(
 	\begin{array}{cc}
 		1 & 0\\
 		0& 1/x
 	\end{array}
 	\right),
 	\label{p2_4}
 \end{equation}
we define
\begin{equation}
T= T_{n(odd)}\circ T_{n(even)}.
\label{p2_9}
\end{equation}
This matrix does not depend on $n$ and it takes the form
\begin{equation}
T= \left(
\begin{array}{cc}
W & Z\\
Z^*& W^*
\end{array}
\right),
\label{p2_10}
\end{equation}
with
\begin{eqnarray}
W&=&\frac{1}{4 x}
\left((1+x)^2 e^{2 i \omega a}-(1-x)^2\right),
\label{p2_11} \\
Z&=&\frac{1-x^2}{4 x}
\left(- e^{2 i \omega a}+1 \right).
\nn\label{p2_12}
\end{eqnarray}
In \Ref{p2_11}, the parameter $a$ is the length of a single section. It is connected with the length $L$ of the string and the number $N$ of the pairs of sections by
\eq{2_13}{ a=\frac{L}{2N}.
}
For a string with non-equal segments one would have to put different $a$'s into the  $Q$'s, \Ref{p1_9}, corresponding to the two matrices entering \Ref{p2_9}. Accordingly, the expressions in \Ref{p2_11} would become more complicated.

The transfer matrix has the property
\begin{equation}
\det T= 1,
\label{p2_13}
\end{equation}
preserving unitarity.
In fact, (\ref{p2_11}) and (\ref{p2_12}) correspond to  Eqs. (13) and (14) in
\cite{brev97-38-2774}.

Now, repeatedly applying \Ref{p2_6}, we get with
\eq{b1a}{\Psi_{2N+1} &= T_{2N}T_{2N-1}\dots T_{1}\Psi_{1}
}	
a relation between the first element of the string and any other element. To obtain the closed string with $2N$ elements we demand periodicity,
\eq{b1}{ \Psi_{2N+1}=\Psi_{1}.
}
At this place, it should be mentioned that a quasi-periodic closure of the string,
\eq{b1q}{ \Psi_{2N+1}=e^{i\Theta}\Psi_{1},
}
results also in real eigenfrequencies. The interpretation could be a charged string, allowed to oscillate only in parallel to a magnetic field  penetrating the loop (other couplings to a magnetic field were considered in \cite{bern97-257-84}). In the antiperiodic case, $\Theta=\pi$, one comes to the twisted string considered in  \cite{bayi96-37-3662}.

With \Ref{b1}, we are faced with a homogeneous system of equations and its determinant must vanish. We define with
\eq{b3}{	\Delta(\omega) \equiv  \det \left(T_{2N}T_{2N-1}\dots T_{1}-{1}\right)
}
the mode generating function. Its zeros, i.e.,   the solutions of the equation
\eq{b4}{\Delta(\omega)=0,
}
are the eigenfrequencies for the vibrations of a closed composite string.
Since we consider a string of equal pairs of sections, all $T_i$ in \Ref{b3} are equal and the mode generating function simplifies to
\begin{equation}
\bigtriangleup(\omega)=\det ( T^N-\mathbbm{1}).
\label{p3_10}
\end{equation}
Now, as a matrix obeys its characteristic equation,  using the property (\ref{p2_13}) and introducing the notation
\begin{equation}
\xi\equiv \frac12 \ \mbox{tr} \, T,
\label{p3_11}
\end{equation}
one arrives at the relation
\begin{equation}
T^2=2\xi T-\mathbbm{1}.
\label{p4_0}
\end{equation}
Repeatedly multiplying the equation (\ref{p4_0}) by $T$ and each time substituting the right-hand side of the equation~(\ref{p4_0}) for $T^2 $, we obtain
\begin{equation}
T^n=T \,u_{n-1}(\xi)-u_{n-2}(\xi).  
\label{p4_1}
\end{equation}
Here, the $u_{n}$ are Chebyshev polynomials,
\begin{equation}
u_{n}(\xi)=\frac{\sin((n+1)\gamma)}{\sin(\gamma)}, \quad \cos (\gamma)=\xi.
\label{p4_2}
\end{equation}
For a chain of delta functions, the approach with Chebychev polynomials was used in \cite{grif01-69-137} for a finite size Kronig-Penney model. Applied there to expressions like \Ref{b1a} with $T_i=T$, these relations  gave closed, explicit expressions for the amplitudes of the wave function. The same would happen for the composite string; however, we do not go into that detail.
We mention, the in \cite{brev97-38-2774} a recursive formula was found, which gives the same results as the application of the Chebyshev polynomials.

Using the above formulas we get from (\ref{p3_10})
\begin{eqnarray}
\bigtriangleup(\omega)&=&\det ( T u_{N-1}(\xi)-(u_{N-2}(\xi)+1)),
\label{p4_3}\\
&=&u_{N-1}(\xi)^2
\nn\\&&-2\xi u_{N-1}(\xi)(u_{N-2}(\xi)+1)+(u_{N-2}(\xi)+1)^2
\nn\label{p4_4},
\end{eqnarray}
for the mode generating function.
After inserting (\ref{p4_2}),   this expression can be simplified  and we obtain
\begin{equation}
\bigtriangleup(\omega)=4 \sin^2\left(\frac{N}{2}\gamma\right) .
\label{p4_6}
\end{equation}
Rewriting the above equation as
$\bigtriangleup(\omega) =  2-\left(e^{iN\gamma}+e^{-iN\gamma}\right)$ and defining	$\alpha=\frac{1-x}{1+x}$,
one can convert the expression (\ref{p4_6}) into the form,
\begin{eqnarray}
\bigtriangleup(\omega) &=& 
2-(1-\alpha^2)^{-N}(\lambda_+^N+\lambda_-^N), 
\\\nn \lambda_{\pm} &=& \cos { (2 a\omega)}-\alpha^2
	\pm\sqrt{(\cos (2a\omega)-\alpha^2)^2-(1-\alpha^2)^2},
\label{p4_6b}
\end{eqnarray}
which was obtained in \cite{brev97-38-2774}.

It should be mentioned \cite{brev97-38-2774}, that the result \Ref{p4_6}, i.e., the expression for the determinant, can be obtained also in an easier way. Let $\lambda_{1,2}$ be the eigenvalues of the transfer matrix $T$,
	\eq{b5}{ \det(T-\lambda_{1,2}\mathbbm{1})=0.
	}
With \Ref{p2_13} we get
\eq{b6}{  \lambda_{1,2}=\xi\pm i\sqrt{1-\xi^2}, \quad \xi=\frac12{\rm tr}T.
}
Now we consider \Ref{p3_10} and diagonalize the matrix, not changing this way its determinant,
\eq{b5a}{	\bigtriangleup
	(\omega)=\det
	\left(
	\begin{array}{cc}\lambda_1^N-1,&0\\0,&\lambda_2^N-1\end{array}
		\right)	=	\left(\lambda_1^N-1\right)\left(\lambda_2^N-1\right)
}
and inserting  $\cos (\gamma)$ for $\xi$ we come also to \Ref{p4_6}.


We use the mode generating function in the form (\ref{p4_6}) as it allows one in the most instructive way  to analyze the structure of the spectrum.
The solutions of (\ref{p4_6})  are
\begin{equation}
 \gamma_j=\frac{2j\pi}{N}, \qquad\left (j ~\mbox{integer}\right).
\label{p4_7}
\end{equation}
Computing  $2\xi$ as the trace of the transfer matrix whith  elements given by  (\ref{p2_10}) and (\ref{p2_11}), we obtain from (\ref{p4_2})
\begin{equation}
\cos (\gamma) =\frac{1}{4x} \left(-(1-x)^2+(1+x)^2 \cos (2a\omega) \right) .
\label{p5_1}
\end{equation}
This is the dispersion relation for the string under consideration. Obviously, $\ga$ is the quasi momentum. To have a real spectrum, the condition
\begin{equation}
-1 \le \frac{ -(1-x)^2+(1+x)^2 \cos (2a\omega) }{4x}\le 1,
\label{p5_2}
\end{equation}
must hold, defining the zone structure. It should be mentioned, that the spectrum is completely discrete (since the closed string has finite spatial extend), and, strictly speaking, there are no bands. This is obvious especially in the case $x=1$ of a completely homogeneous string having an equidistant spectrum. However, with more sections on the string, the eigenvalues start to group; forming zones eventually when their number reaches infinity.

\begin{figure}[h]
	\centering
	\includegraphics[width=0.95\linewidth]{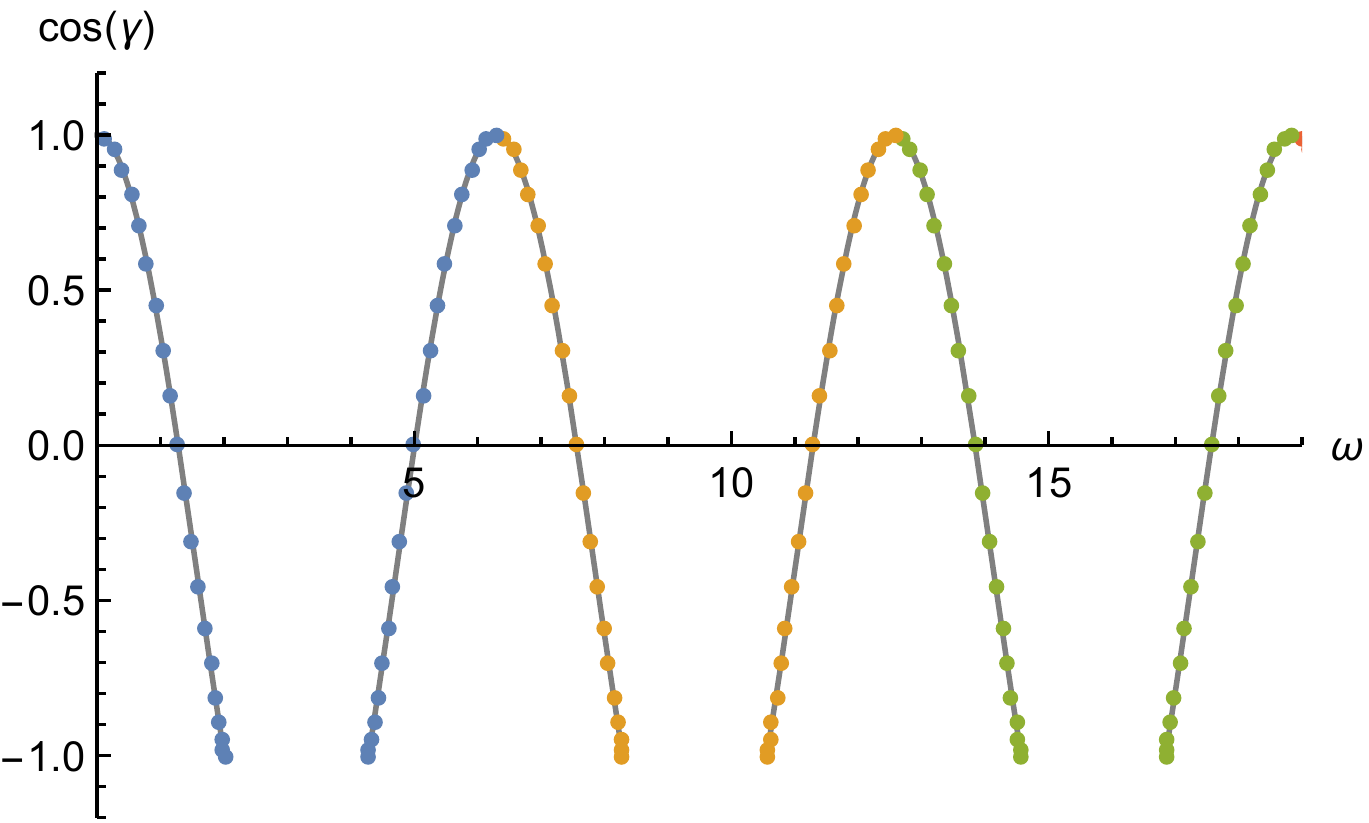}
	\caption[Spectrum]{The spectrum according to eq. \Ref{p5_1} for a string with $N=40$ sections, $x=0.3$ and  $L=1$. The dots denote the eigenvalues. For $N\to\infty$ or $L\to\infty$ these approach the solid line.}
	\label{fig:1}
\end{figure}

Eq. \Ref{p5_1} can be inverted easily,
\eq{p5_3}{\cos (2a\omega)=\frac{4 x \cos (\gamma)+(1-x)^2}{(1+x)^2},
}
and with $\ga_j$, \Ref{p4_7}, we get  explicit expressions,
\eq{p5_4}{
	\omega_{n,j}
	&=\frac{\om_j+2\pi n}{2a},
		~~~\left(j=1,\dots,\left[\frac{N}{2}\right],~~n=0,1,\dots\right),
\nn\\
\omega_{n,j}
&=\frac{2\pi-\om_j+2\pi n}{2a},
\\&\nn	~~~\left(j=0,\dots,\left[\frac{N}{2}\right],~~n=0,1,\dots\right),,
}
where we defined
\eq{b52}{ \om_j=\arccos \left(\frac{4 x \cos (\gamma_j)+(1-x)^2}{(1+x)^2}\right),
}
for the eigenfrequencies. The $\arccos$ is defined on its main branch. The constant mode, i.e., $\om=0$, is excluded since it does not contribute to the energy. An example is shown   in Fig. \ref{fig:1}.
The degeneracy of the modes results in a weight function
\eq{3.3}{ \chi_j=\left\{\begin{array}{l}
		1~~~\mbox{for}~N~\mbox{even and }~j=\frac{N}{2},
		\\   2~~~\mbox{otherwise}.  \end{array}\right.
}
The degeneracy is 2 except for the modes with $j=\frac{N}{2}$ for even $N$.
%

\section{\label{T3}Vacuum energy}
In zeta functional regularization,  the  vacuum energy  is defined as
\begin{equation}
E_0=\frac{\mu^{2s}}{2}\sum_{(n)}\omega_{(n)}^{1-2s}.
\label{p6_1}
\end{equation}
The sum goes over all nonzero eigenvalues $\omega_{(n)}$ with the corresponding multiplicity. In our case, the eigenvalues are specified by \Ref{p5_4}.
The arbitrary parameter $\mu$ comes in with the regularization and is chosen as to preserve the correct dimension of the regularized energy.

It should be mentioned that frequently the vacuum energy is considered relative to the vacuum energy of the empty space. Accordingly, in all papers on the composite string, it is considered relative to that of the homogeneous string. To be more precise,  people consider the difference between the vacuum energy of the composite string and that of the homogeneous string. This procedure is meaningful in the case of an infinite space since its vacuum energy is proportional to its (infinite) volume. In the case of the closed string, which has with its finite length a finite 'volume', we consider the mentioned subtraction as superfluous and consider the complete vacuum energy.

There are two (main) approaches to calculate the vacuum energy. One is to transform the sum in \Ref{p6_1} into an integral and to move the integration path towards the imaginary axis. This way is especially preferable, not only by avoiding to work with oscillating quantities as one has typically for real frequencies but for easy separation of the volume contribution for problems in infinite volume. Also, it allows for an easy generalization to Matsubara representation (at finite temperature). The other way is a direct summation in  \Ref{p6_1}. This is especially preferable for a linear spectrum, allowing easily for results in terms of the Riemann zeta function or related zeta functions. Many examples of this kind are collected in \cite{eliz95b}. Below, we discuss both approaches.

\subsection{\label{T31}Vacuum energy in integral approach}
In this approach one starts from the frequencies $\omega_{n,j}$ as  solutions of the equation \Ref{b4}. With
\eq{h1}{ \Delta(\om)=4 g(\om)^2,~~~g(\om)=\sin\left(\frac{N}{2}\ga\right),
}
where $\ga$ is given by \Ref{p5_1}, we define a modified mode generating function $g(\om)$  having   zeros in the same locations as \Ref{b4}, but as single zeros.

We are going to transform the sum in \Ref{p6_1} into a contour integral. For that, we define with
\eq{p6_2}{ h(\om)=\ln\frac{g(\om)}{g_0(\om)}, ~~~g_0(\om)=\frac{N\om(1+x)}{4\sqrt{x}},
}
a function whose logarithmic derivative has  single zeros in the eigenfrequencies, $\om_{n,j}$, and which vanishes at the origin,
\eq{p8_2}{h(\om) {\underset{\om\to0}{\sim}}\om^2.
}
With these,  the vacuum energy becomes	
\eq{p6_3}{ E_0=  {\mu^{2s}} \int\limits_{C}^{}\frac{d\omega}{2\pi i}\omega^{1-2s}\partial_{\omega}h(\omega),
}
where we accounted for a factor of 2 which results from switching from the mode generating function $\Delta(\om)$ to $g(\om)$ in eq. \Ref{h1}.
The contour $C$ encircles the real $\om$-axis. We mention that we excluded any contribution from the origin by dividing by $g_0(\om)$ in \Ref{p6_2}. This way we exclude the constant mode which is in this case with a massless field a zero-mode and which should not enter in zeta-functional regularization.

We continue with the Wick rotation, $\omega=i\zeta$, on the upper half of the integration path $C$,  and $\omega=-i\zeta$ on the lower half. We get with
\eq{p7_2}{
	E_0=-\frac{\cos(\pi s)}{\pi} \mu^{2s}\int\limits_{0}^{\infty}
	d\zeta \, \zeta^{1-2s} \partial_{\zeta}h(i\zeta),
}
a representation in terms of imaginary frequency $\zeta$, which is, as usual, convenient for further work. The function $h(i\zeta)$ has a quite simple explicit form,
\begin{widetext}
\eq{hsub2}{ h(i\zeta) = \ln \left( \sinh\left(\frac{N}{2}{\rm arccosh}\frac{-(1-x)^2+(1+x)^2\cosh(2a\zeta)}{4x}\right)\right)
	-\ln\left(\frac{N2a\zeta(1+x)}{4\sqrt{x}}\right),
}
\end{widetext}
which is in terms of real functions.

We mention the asymptotic properties
\eq{as1}{
	h(i\zeta){\underset{\zeta \to 0}{\sim}\zeta^2},
	~~~~h(i\zeta){\underset{\zeta \to \infty}{\sim}\zeta}
}
allowing the integration in \Ref{p7_2} to converge for $1<s<\frac32$.

We have to construct the analytic continuation to $s=0$. For this, we define  functions
\eq{has}{	h^{as}&=\frac{h^{inf}}{1+\zeta^{-3}},
\\\nn	h^{inf}&= \frac{N \zeta}{2}+(N-1)\ln\left(\frac{(1+x)}{2\sqrt{x}}\right) -\ln\left(\frac{N\zeta}{2}\right),
\\\nn  h^{sub}&=h-h^{as}
}
having the properties
\eq{hsub1}{	h^{sub} \underset{\zeta \to 0}{\sim}{\zeta^2},
		~~~	h^{sub} \underset{\zeta \to \infty}{\sim}{\zeta^{-2}},
\\\nn		h^{as} \underset{\zeta \to 0}{\sim}{\zeta^{-2}},
	~~~	h^{as} \underset{\zeta \to \infty}{\sim}{\zeta}.
}
Here we dropped for  a moment the arguments of the functions to simplify notations.
The function $h^{inf}$ is just the asymptotic expansion of $h$, \Ref{hsub2}, for $\zeta\to\infty$ up to exponentially decreasing terms.
We split the vacuum energy according to
\eq{E01}{E_0=& E^{fin}+E^{as},
}
into finite and asymptotic parts with
\eq{E02}{ E^{fin}&= \frac{1}{\pi}\int_0^\infty d\zeta \ h^{sub},
\\\nn E^{as}&= \frac{(1-2s)\cos(\pi s)}{\pi}\mu^{2s}\int_0^\infty d\zeta \ \zeta^{-2s}h^{as}.
}
In $E^{fin}$ we could remove the regularization by putting $s=0$ due to the decrease \Ref{hsub1}. Also,  we integrated by parts which is possible without surface terms, also due to \Ref{hsub1}. This expression has to be evaluated numerically.

\begin{figure}[t]
	\centering
	\includegraphics[width=0.95\linewidth]{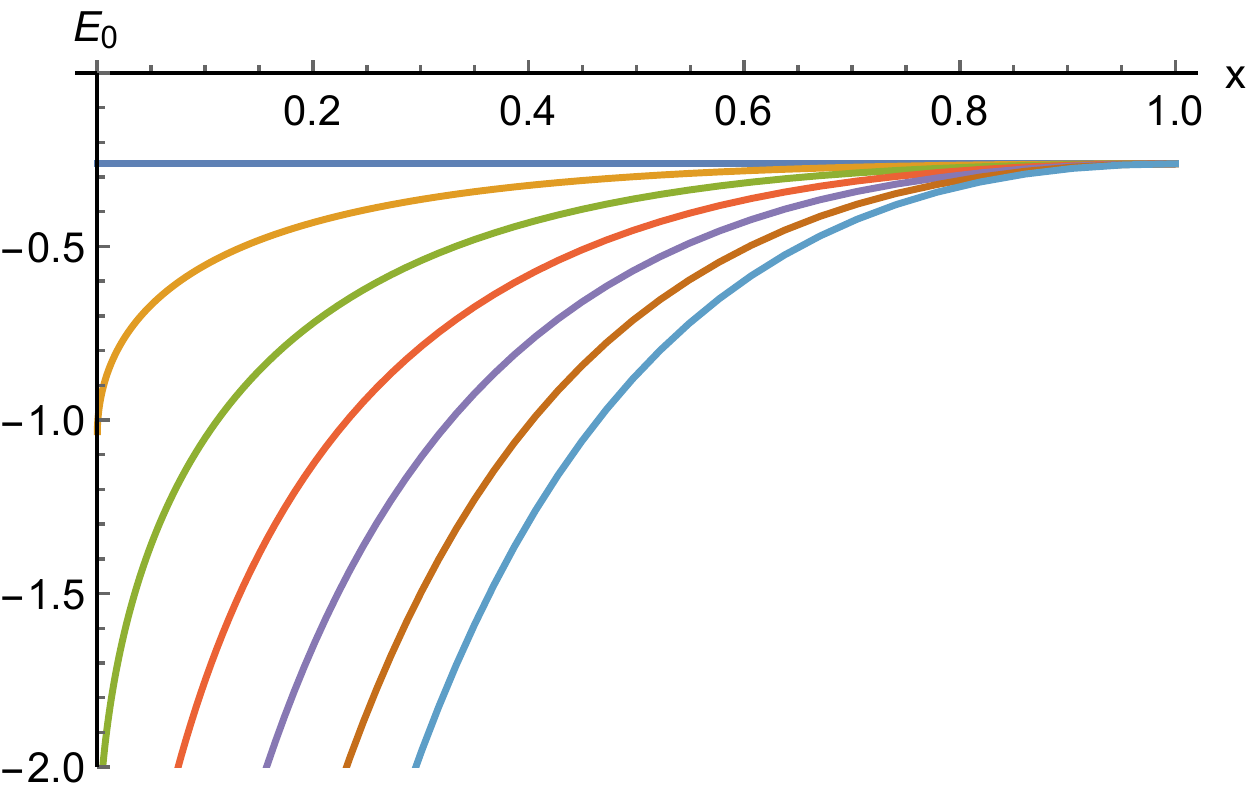}
	\caption[E0]{The vacuum energy $E_0$, \Ref{E01}, \Ref{E2}, for a composite string as a function of the ratio $x$, \Ref{2_3}, of the string tensions for several numbers $N$ of sections of the string. $N$ increases from top to bottom (color online). The length of the string is $L=1$.}
	\label{fig:2}
\end{figure}

The asymptotic part, $E^{as}$ in (\ref{E02}), can be integrated explicitly with the result
\begin{widetext}
\eq{Eas1}{
		E^{as}&=\frac{(\mu L/N)^{2s}}{6}  \frac{  (2 s-1) \cos(\pi s)}{\sin \left(\frac{\pi}{3} (2   s+1 )\right) \cos^2\left(\frac{\pi}{6} (4   s+1 )\right) } \\\nonumber
		&~~~~~~\times\left(\left(2 \cos \left(\frac{4 \pi  s}{3}\right)+1\right) [(\text{N}-1) \ln \gamma_0-\ln (\text{N})]+ \text{N} \cos \left(\frac{\pi}{3} (4   s+1 )\right)+ \text{N}+\frac{2\pi}{3}   \sin \left(\frac{4 \pi  s}{3}\right)+\frac{\pi}{\sqrt{3}}\right).
}
\end{widetext}
The analytic continuation to $s=0$ reveals no pole.
This way, within the zeta functional regularization there is no ultraviolet divergence.  This is in accordance with the heat kernel coefficients which will be discussed in Sect.\ref{T43}.
Finally, for $s=0$, the asymptotic part  reads
\eq{Eas2}{
	E^{as}=-\frac{4}{3\sqrt{3}} \,\frac{N}{L}\left( \left((N-1)\ln \gamma_0- \ln N \right)
	+  \frac{N}{2}+ \frac{ \pi}{3\sqrt{3}}\right).
}
To compute the vacuum energy we scale out the dimensional parameters, $\zeta\to\frac{\zeta}{2a}$, to rewrite  the finite part of the energy, \Ref{E02}, in the form
\eq{E2}{E^{fin} &= \frac{N}{\pi L}\int_0^\infty d\zeta \ h^{sub}\left(i\zeta\frac{L}{N}\right),
}
where the integral is now dimensionless. Examples for $E_0$ are shown in Fig. \ref{fig:2}.

\subsection{\label{T32}Vacuum energy in sum approach}
The sum representation \Ref{p6_1} of the vacuum energy needs to be specified in terms of the eigenfrequencies \Ref{p5_4}. Accounting for the multiplicity \Ref{3.3}, it reads
%
%
\eq{3.2}{ E_0 &= \frac{1}{2(2a)^{1-2s}}
	\left[           \sum_{j=1}^{\left[\frac{N}{2}\right]}
	\sum_{n=0}^{\infty}  \chi_j  \left(\om_j+2\pi n\right)^{1-2s}
\right.\\\nn		&\left.	+ \sum_{j=0}^{\left[\frac{N}{2}\right]}
				\sum_{n=0}^{\infty}  \chi_j  \left(2\pi-\om_j+2\pi n\right)^{1-2s}
			\right].
}
It should be mentioned, that the account for the multiplicity, which here is taken care of by the function $\chi_j$, happens automatically in the integral approach.

The sums over $n$ define Hurwitz zeta functions, as first in this context mentioned in \cite{lixi91-44-560}, and we arrive at
\eq{3.4}{	E_0 &= \frac12  \left(\frac{\pi}{a}\right)^{1-2s}	\left[
	\sum_{j=1}^{\left[\frac{N}{2}\right]} \chi_j\,
		\zeta_{\rm H}\left(2s-1, \frac{\om_j}{2\pi} \right )
\right.\\\nn&\left.~~~~~~	+ \sum_{j=0}^{\left[\frac{N}{2}\right]} \chi_j\,
	\zeta_{\rm H}\left(2s-1,1- \frac{\om_j}{2\pi} \right)  \right].
}
Now the analytic continuation in $s$ is given by the properties of the Hurwitz zeta function and we arrive immediately at
\eq{3.5}{	E_0 &= \frac{\pi}{2a}
	\left[\sum_{j=1}^{\left[\frac{N}{2}\right]} \chi_j\,
	\zeta_{\rm H}\left(-1, \frac{\om_j}{2\pi}\right )
	+ \sum_{j=0}^{\left[\frac{N}{2}\right]} \chi_j\,
	\zeta_{\rm H}\left(-1,1- \frac{\om_j}{2\pi}\right ) \right].
}
Again, as in the preceding section, we observe no pole in $s$.
We mention the known relation $\zeta_{\rm H}(-1,1-a)=\zeta_{\rm H}(-1,a) $, allowing for some insignificant simplification. Further, we mention that the Hurwitz zeta function with negative integer argument has an explicit expression in terms of Bernoulli polynomials. Specifically in our case
\eq{3.6}{\zeta_{\rm H}(-1,a) =-\frac12\left(\frac16-a+a^2\right)
}
holds, showing that we have with  \Ref{3.5}  an expression for the vacuum energy of the composite string in terms of finite sums over polynomials.
This representation is, of course, equivalent to \Ref{3.5} and \Ref{E01} in the sense that these and \Ref{3.5} are different representations of the same quantity. The sum representation \Ref{3.5} appears to be somehow simpler.

\section{\label{T4}Limiting cases of the vacuum energy and the heat kernel coefficients}
%
\subsection{\label{T41}The cases $x=0$ and $x=1$}
The limiting cases $x=0$ and $x=1$ (homogeneous string) can be easiest obtained from the sum representation \Ref{3.5}. For $x=0$ we mention with \Ref{b52} $\om_j=0$ and with $\zeta_{\rm H}(-1,0)=-\frac{1}{12}$ we get a sum over the $j$'s in the form
\eq{3.7}{	\sum_{j=1}^{\left[\frac{N}{2}\right]} \chi_j\,
	+\sum_{j=0}^{\left[\frac{N}{2}\right]} \chi_j\,=      2N.
}
For $x=1$ we note $\om_j=\ga_j$ and use the property
\eq{3.8}{\sum_{k=1}^n\zeta_{\rm H}\left(-1,\frac{k}{n}\right)
	=-\frac{1}{12n}
}
of the Hurwitz zeta function. The sums in \Ref{3.5} collect just into the form \Ref{3.8}. As a result, for this two cases we get
\eq{3.9}{	{E_0}_{|x=0} &= -\frac{\pi N^2}{6L},
~~~	{E_0}_{|x=1} = -\frac{\pi }{6L}.
}
The same results can be obtained from the integral approach.

\subsection{\label{T42}Limiting cases $N\to\infty$}
We consider this limit in both representations and start with the integral representation, Sect. \ref{T31}. First, we consider $E^{fin}$, \Ref{E02}. It is possible to perform the limit $N\to\infty$ under the sign of the integral. We get from \Ref{hsub2} and \Ref{has}
\eq{4.1}{h^{sub}   & \underset{N\to\infty}{\sim} h^{sub}_{inf}
\\\nn
h^{sub}_{inf} &=
	\frac{N}{2}{\rm arccosh}\frac{-(1-x)^2+(1+x)^2\cosh(\zeta)}{4x}
\\\nn&~~~~	-\frac{1}{1+\zeta^{-3}}\left(\frac{N\zeta}{2}+N\ln\left(\frac{1+x}{2\sqrt{x}}\right)\right).
}
We denote the finite part of the vacuum energy in this limit by $E^{fin}_{inf}$ and with \Ref{4.1}{ it has the representation
\eq{4.2}{ E^{fin}_{inf} = \frac{1}{\pi}\int_0^\infty d\zeta\ h^{sub}_{inf}.
}
From the asymptotic part, \Ref{Eas2}, we have in this limit
\eq{4.3}{	 E^{as}_{inf}  = \frac{4N}{3\sqrt{3}}
	\left( \ln\left(\frac{1+x}{2\sqrt{x}}\right) +\frac12\right)
}
and together we get
\eq{4.4}{ E_0 \underset{N\to\infty}{\simeq}  E^{fin}_{inf}+ E^{as}_{inf}.
}
which is proportional to $N$, and after restoring the dimensions to $N^2$.

Another approach starts with the sum representation \Ref{3.5}. In the limit $N\to\infty$, the sum over $j$ turns into an integration according to the rules
\eq{4.5}{ \frac{j}{N} \to \frac{\ga}{2\pi},
	~~~\sum_{j=0}^{\left[\frac{N}{2}\right]} \to \frac{N}{2\pi}\int_0^\pi d\ga.
}
In this limit, the differences between the starting points in the two sums in \Ref{3.5}, as well as the factor $\chi_j$, become unimportant and we end up with
\eq{4.6}{  E_0 \underset{N\to\infty}{\simeq} \frac{ 2 N^2}{L} w(x),
	~~~w(x)= \int_0^\pi d\ga\ \zeta_{\rm H}\left(-1,\frac{\om(\ga)}{2\pi}\right),
}
with
\eq{4.7}{\om(\ga) = \arccos\frac{4x\cos(\ga)+(1-x)^2}{(1+x)^2}.
}
Both expressions, \Ref{4.4} and \Ref{4.6}, represent the same
limit of the vacuum energy. The first one involves   infinite integration, the other a finite one. Both integrations cannot be done analytically; however, it is easy to evaluate them numerically. A plot is shown in Fig. \ref{fig:3}. In this plot, for $x=0$, we note from \Ref{4.7} $\om(\ga)=0$ and we get $E_0\sim -N^2/6L$. In $x=1$ we have $\om(\ga)=\ga$ and the integration over $\ga$ gives zero.

\begin{figure}[t]
	\centering
	\includegraphics[width=0.95\linewidth]{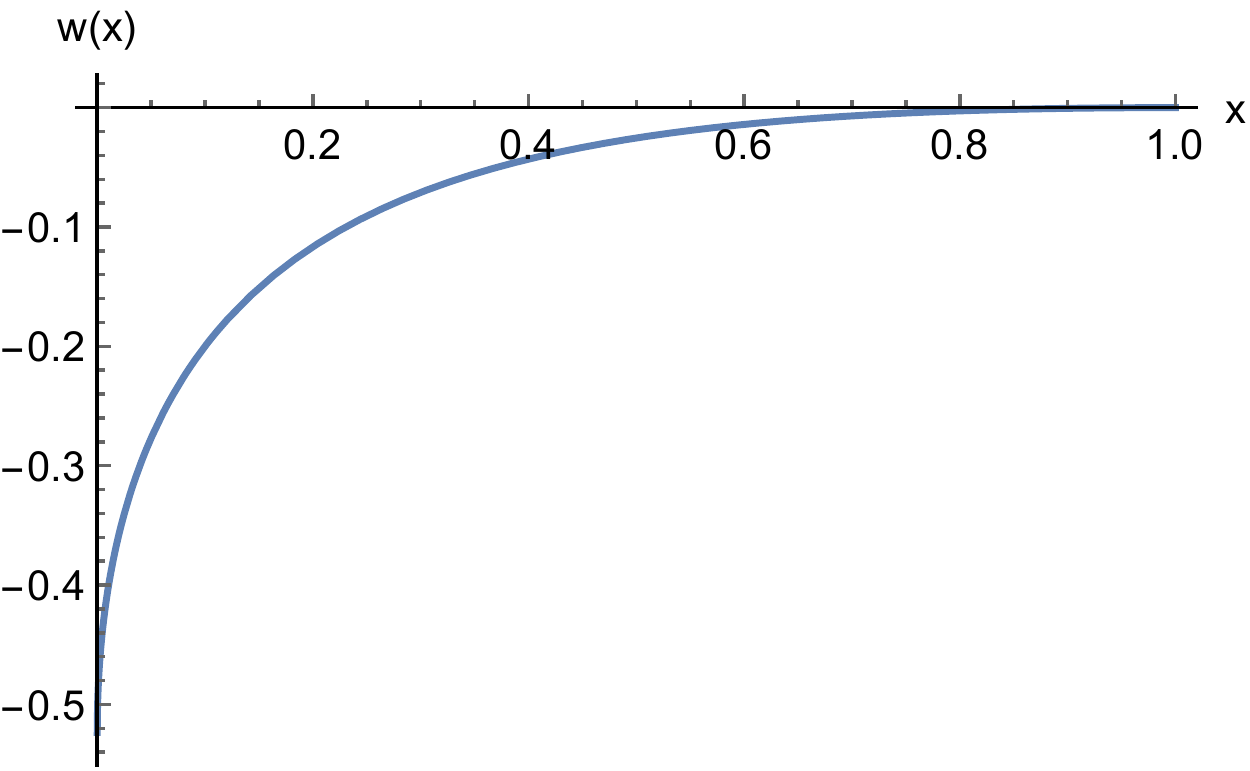}
	\caption[Spectrum]{The limiting slope $w(x)$ of the vacuum energy\Ref{4.6}  for $N\to\infty$ as a function of the ratio $x$.}
	\label{fig:3}
\end{figure}

\subsection{\label{T43}The heat kernel coefficients}
The heat kernel expansion and its coefficients are the universal tool to investigate the ultraviolet behavior of the vacuum energy. At once this is a semiclassical expansion since in powers of $\hbar$ and, at once, in inverse powers of the mass (if present). In terms of eigenvalues, the heat kernel and its expansion are  defined as,
\eq{4.8}{ K(t)=\sum_{(n)}e^{-t\om_{(n)}^2}
	\underset{t\to 0}{\simeq}  \frac{1}{(4\pi t)^{d/2}}\sum_{n\ge 0}a_nt^n,
}
where $d$ is the space dimension, $d=1$ in our case, and the $a_n$ are the heat kernel coefficients. To calculate the coefficients we use the relation to the zeta function $\zeta_{\mathbf{P}}(s)$, \Ref{A1},  of the composite string. Using an integral representation we get
\eq{4.9}{ \zeta_{\mathbf{P}}(s) &=
	\int_0^\infty\frac{dt}{t}\,  \frac{t^s}{\Gamma(s)}
	\sum_{(n)}e^{-t\om_{(n)}^2}
	=	\int_0^\infty\frac{dt}{t}\,  \frac{t^s}{\Gamma(s)} K(t).
}
The behavior of the integrand for small $t$ results in poles which can be determined by inserting the expansion \Ref{4.8} into \Ref{4.9}. Integrating $t$ from $0$ to $1$ one gets  the pole part,
\eq{4.10}{ \zeta_{\mathbf{P}}(s) &= \simeq
	\sum_{n\ge 0}\frac{a_n}{\sqrt{4\pi}\,\Gamma(s)} \ \frac{1}{s+n-\frac12}
+\dots,
}
and the dots denote the regular part. This formula allows   calculating the coefficients from the residua,
\eq{4.11}{  a_n=\Res_{s=\frac12-n}\sqrt{4\pi}\,\Gamma(s) \zeta_{\mathbf{P}}(s) .
}
As a special case we mentions that for $n=\frac12$ the relation
\eq{4.13}{ a_\frac12 = 2\sqrt{\pi}\zeta_{\mathbf{P}}(0)
}
follows.

Next, we use the information on $\zeta_{\mathbf{P}}(s)$, which is collected in the Appendix, eq. \Ref{A5}. For $n=0$, we have the pole of $\zeta_{\mathbf{P}}(s)$ in $s=\frac12$.
 For $n=\frac12$ we have the pole of the gamma function in $s=0$ and for $n=1$ we have no pole. Accordingly, the coefficient is zero, which is in agreement with the absence of ultraviolet divergences observed earlier. With there remarks, the coefficients become
\eq{4.12}{ a_0 &= L,
	\\\nn a_\frac12  &= -\sqrt{\pi},\\\nn a_1 &=0.
}
All higher order coefficients are zero. This is in agreement with the observation that there are only exponentially small corrections to the asymptotic expansion $h^{inf}$, defined in \Ref{has}, of the mode generating function $h$, \Ref{hsub2}, in \Ref{p7_2}.

It should be mentioned that the coefficient $a_\frac12$, which is non-zero, does not depend on the parameters of the string. For instance, it does not disappear when taking the limit of the homogeneous string ($x\to 1$). Similar features were observed in \cite{bord05-38-11027} for a flat plasma sheet in the TM-mode and, similar,  in \cite{bord08-77-085026} for a spherical plasma shell. As discussed in \cite{bord18-51-455001} (Sect. 4), this is related to the Klauder-phenomenon stating that some singular perturbations (like the jump between the sections of the string in our case) cannot be turned off to restore the unperturbed situation. Doubts in the physical meaning of such a situation were discussed in \cite{milt19-99-045013}.

\section{\label{T5}The free energy}
At finite temperature, there are basically two approaches. One is in  terms of the Matsubara frequencies, the other in terms of real frequencies. In the first one, which was used also in the literature (\cite{brev95-51-1869}, \cite{brev98-15-3383}, \cite{brev03-44-1044})), one starts from the integral representation in terms of imaginary frequencies like \Ref{p7_2} after integrating by parts,
\eq{5.1}{	E_0=\frac{(1-2s)\cos(\pi s)}{\pi} \mu^{2s}\int\limits_{0}^{\infty}
	d\zeta \, \zeta^{-2s} \,h(i\zeta),
}
and substitutes the integration by a sum,
\eq{5.1a}{ \int_0^\infty d\zeta f(\zeta) \to T
	{\sum_{l=0}^{\infty}{\vphantom{\sum}}^{\prime}}
	 f(2\pi T l),
}
(the contribution from $l=0$ enters with weight $\frac12$). As mentioned, this expression contains the ultraviolet divergence. In the given case, the simplest way to get rid of it is to subtract the homogeneous string contribution and to treat it separately. After that, one can put $s=0$ and comes to the conventional shape of this representation.

The other representation is
\eq{5.2}{ F=\sum_{n,j}\left(\frac{\om_{n,j}}{2}+T\ln\left(1-e^{-\om_{n,j}/T}\right)\right).
}
The first term in the parenthesis is the vacuum energy and the second is the temperature-dependent part, $\Delta_{\rm T}F$, of the free energy. In the following, we focus on it and on the entropy $S=-\frac{\pa F}{\pa T}$. We represent these in the form
\eq{5.3}{ \Delta_{\rm T}F=T\sum_{n,j}f\left(\frac{\om_{n,j}}{T}\right),
	~~~ S=\sum_{n,j}s\left(\frac{\om_{n,j}}{T}\right),
}
with
\eq{5.4}{ f(\om)=\ln\left(1-e^{-\om}\right),
	~~~g(\om)= -\ln\left(1-e^{-\om}\right)+\frac{\om}{e^\om-1}.
}
With the summations defined as in \Ref{3.2}, since these sums are fast converging now, one can easily produce numbers and plots. Examples are shown in Fig. \ref{fig:4}. As can be seen, the free energy is monotone and so is the entropy. It has the right sign and vanished at the origin. Thus, the thermodynamics of the considered system does not bear an interesting feature like that mentioned in the Introduction.


\onecolumngrid

\begin{figure}[h]
	\centering
	\includegraphics[width=0.475\linewidth]{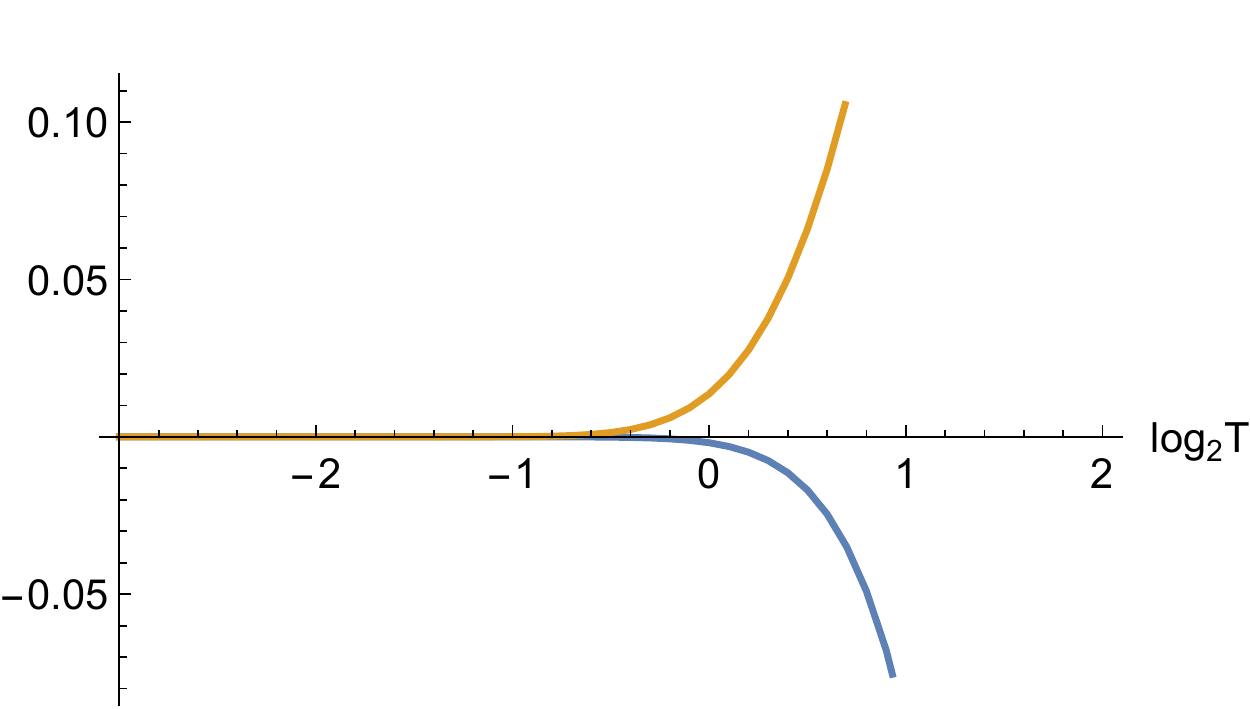}
	\includegraphics[width=0.475\linewidth]{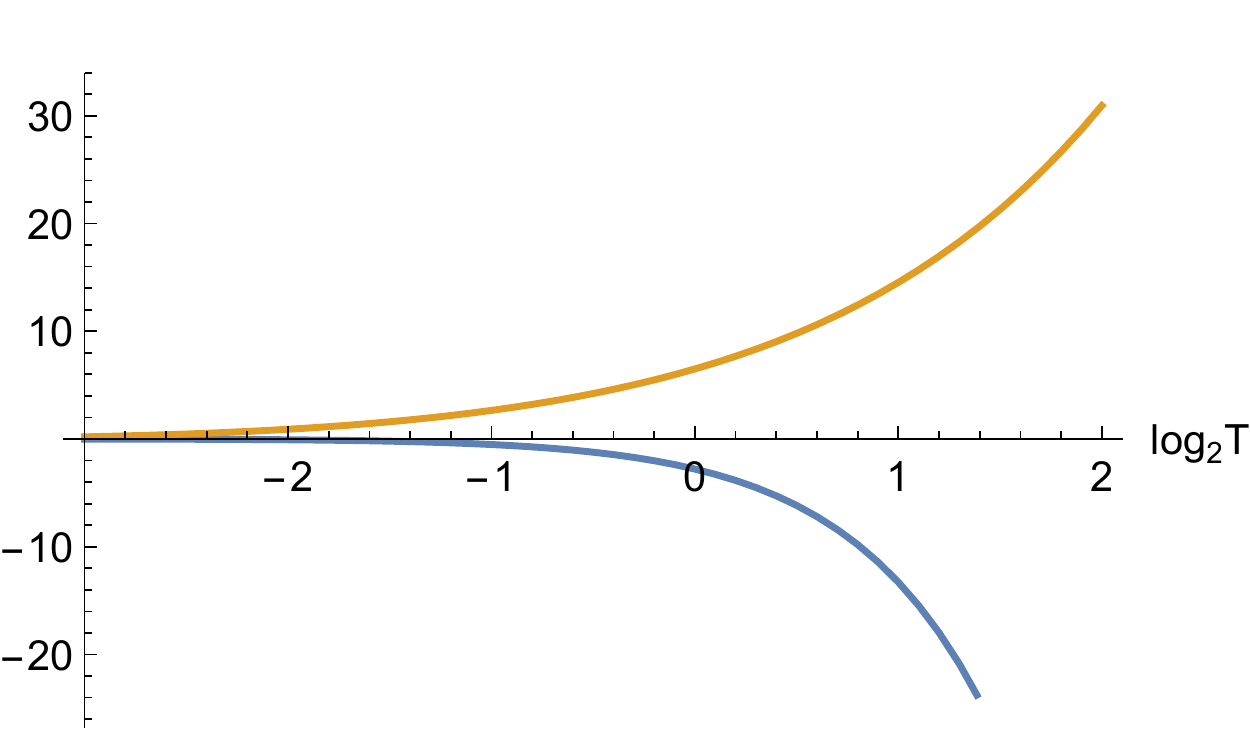}
	\caption[Free energy]{The free energy (lower curve) and the enropy (upper curve)  of the composite string for $N=1$ (left panel) and $N=16$, $x=0.9$ (right panel).
	}
	\label{fig:4}
\end{figure}

\twocolumngrid

The limiting cases for small and high temperatures can be obtained easily, following Chapt. 5 in \cite{BKMM}, for instance. The low-temperature behavior, since the spectrum is discrete, is determined by the lowest non zero eigenvalue,
$\om_{0,1}=\frac{1}{2a}\arccos \left(\frac{4 x \cos (\frac{2\pi}{N})+(1-x)^2}{(1+x)^2}\right)$, in \Ref{p5_4}, to be
\eq{5.5}{\Delta_{\rm T}F \underset{T\to0}{\simeq}T\left(1-e^{-\om_{0,1}/T}\right).
}
For the high-temperature expansion we use the heat kernel expansion. The ready-to use formula (5.53) in \cite{BKMM} is for (3+1)-dimensions. Therefore we go back to (5.43),
\eq{5.6}{ F &= -\frac{T}{2}\pa_s\mu^{2s}
	\int_0^\infty\frac{dt}{t}\,\frac{t^s}{\Gamma(s)}
	\sum_l e^{-t\xi_l^2} K(t),
}
where $\xi_l=2\pi T l$ are the Matsubara frequencies, and we have to put $s=0$ at the end.
This is an expression of  the free energy in terms of the heat kernel. The high-$T$ expansion   follows from the heat kernel expansion. Inserting \Ref{4.8} with $d=1$, and separating the contribution from $l=0$, we arrive at
\begin{widetext}
\eq{5.7}{ F\simeq
	-\frac{T}{2}\pa_s\mu^{2s}
	\left[ \zeta_{\mathbf{P}}(s)
	+\sum_{n\ge0} a_n 	\int_0^\infty\frac{dt}{t}\,\frac{t^{s+n-\frac12}}{\sqrt{4\pi}\Gamma(s)}
	2\sum_{l=1}^\infty e^{-t\xi_l^2}
\right].
}
Now the integration over $t$ can be carried out. Subsequently, the sum over $l$ gives a Riemann zeta function and we arrive at
\eq{5.8a}{ F\simeq
	-\frac{T}{2}\pa_s\mu^{2s}
	\left[ \zeta_{\mathbf{P}}(s)
	+\sum_{n\ge0} a_n 	\frac{\Gamma(s+n-\frac12)}{\sqrt{4\pi}\,\Gamma(s)}
		\,	(2\pi T)^{1-2s-2n}2\zeta_{\rm R}(2s+2n-1)
	\right],
}
which for $d=1$ comes in place of (5.49) in \cite{BKMM}.
Now we  take the derivative, put $s=0$ and arrive at the high-$T$ expansion in the form
\eq{5.8}{ F \underset{T\to\infty}{\simeq}
-	\frac{\pi}{6}a_0T^2
	-\left(\frac{\ln(T)}{2\sqrt{\pi}}a_\frac12+\frac12\zeta'_{\mathbf{P}}(0)\right)T
	+\frac{1}{4\pi}\left(\ln\frac{4\pi T}{\mu}-\gamma\right)a_1
	+O(T^{-1}).
}
\end{widetext}
Inserting the coefficients \Ref{4.12} we arrive at the expansion
\eq{5.9}{	 F \underset{T\to\infty}{\simeq}
-	\frac{\pi L}{6}T^2+\frac12\left(\ln(T)- \zeta'_{\mathbf{P}}(0)
	\right)T+\dots\,.
}
The corrections are exponentially small like in the case of the Casimir free energy for ideal parallel plates in accordance with the vanishing of the higher-order heat kernel coefficients.
The first term is   black-body radiation in the given case.

\section{\label{T6}Conclusions}
The composite string is non-trivial, but simple model
for studying  vacuum (Casimir) energy and
thermodynamic properties. It allows us to demonstrate the basic technical tools and to get the most explicit results. The model itself is not very interesting; however easy generalizations may reveal more exquisite features like Hagedorn temperature or instabilities,  which were discussed in the literature.

In the present paper, we recalculated the mentioned quantities in two representations. We used only zeta functional regularization well knowing that all other regularizations will be equivalent. The system has a vanishing heat kernel coefficient $a_1$. Thus, there are no ambiguities in its renormalization and in zeta-functional regularization no renormalization is needed.
The thermodynamic properties are most simple, free energy and entropy are monotone functions; in opposite to some mentioned other simple systems.

The authors  hope that the above presentation of the topic may serve as a good starting point for more interesting applications and further developments.
\\ \\

\begin{acknowledgments}
	This paper is an extended version of a talk one of us (IGP) gave on  {\it The 10th International workshop
		“Waves in inhomogeneous media and integrable systems”}, held in
	September 24-25, 2020, at IKBFU,
	Kaliningrad, RF, and we thank the organizers for the opportunity to present this talk.
\end{acknowledgments}

\appendix*
\section{\label{A}The zeta function of the composite string}
The zeta function $\zeta_{\mathbf{P}}(s)$ of the composite string is the zeta function associated with the operator $\mathbf{P}$ which is determined by the equation \Ref{1.1} and the matching conditions \Ref{1.1a} and \Ref{b1}. In terms of its eigenvalues, $\om_{(n)}$,  it is given by
\eq{A1}{ \zeta_{\mathbf{P}}(s) = \sum_{(n)}{\om_{(n)}}^{-2s}.
}
This is similar to the definition \Ref{p6_1} of the vacuum energy and the relation
\eq{A2}{ E_0(s)=\frac{\mu^{2s}}{2}\,\zeta_{\mathbf{P}}(s-\frac12)
}
holds. Using the same steps which resulted in eq. \Ref{3.4}, we get
\begin{widetext}
\eq{A3}{\zeta_{\mathbf{P}}(s) &= \left(\frac{\pi}{a}\right)^{-2s}
	\left(
	\sum_{j=1}^{\left[\frac{N}{2}\right]} \chi_j\,
	\zeta_{\rm H}\left(2s, \frac{\om_j}{2\pi} \right )
	+ \sum_{j=0}^{\left[\frac{N}{2}\right]} \chi_j\,
	\zeta_{\rm H}\left(2s,1- \frac{\om_j}{2\pi} \right)    \right).
}
\end{widetext}
To continue we use the following properties of the Hurwitz zeta function,
\eq{A4}{	\zeta_{\rm H}(2s,a) &=\frac{1}{2s-1}+\dots ~~~\mbox{for}~~s\to\frac12,
\\\nn		\zeta_{\rm H}(0,a) &=\frac12-a
\\\nn	\zeta_{\rm H}(-1,a) &=-\frac12\left(\frac16-a+a^2\right).
}
Further, we need to know that $	\zeta_{\rm H}'(0,a) $ (the derivative with respect to $s$) is a finite function of $x$ and $N$.

With these properties, it is easy to get the following relations,
\eq{A5}{	
	\zeta_{\mathbf{P}}(s) &= \frac{N}{2\pi}\frac{1}{s-\frac12}+\dots ~~~\mbox{for}~~s\to\frac12,
\\\nn	\zeta_{\mathbf{P}}(0) &= -\frac{\pi N}{L}.
}
In the first one we used
\eq{A6}{ \sum_{j=1}^{\left[\frac{N}{2}\right]} \chi_j\,
	+ \sum_{j=0}^{\left[\frac{N}{2}\right]} \chi_j\, =2N	
}
and for the second one, with the second line in \Ref{A4}, we note
\eq{A7}{  \sum_{j=1}^{\left[\frac{N}{2}\right]} \chi_j\, \left(\frac12-\frac{\om_j}{2\pi}\right)
	+ \sum_{j=0}^{\left[\frac{N}{2}\right]} \chi_j\,\left(-\frac12+\frac{\om_j}{2\pi}\right)
	=-1.
}
Here all terms except the first one ($j=0$) canceled and $\om_0=0$ holds. Finally, we mention that the value of $	\zeta_{\mathbf{P}}(s)$ at $s=-\frac12$ is the vacuum energy, \Ref{A2}, and the derivative in $s=-\frac12$ is a finite function.


\begin{thebibliography}{10}

\bibitem{brev90-41-1185}
I~Brevik and H~B Nielsen.
\newblock {Casimir Energy for a Piecewise Uniform String}.
\newblock {\em Phys.~Rev.~D}, {41}({4}):{1185--1192}, {1990}.

\bibitem{brev99-453-217}
I.~Brevik, A.A. Bytsenko, and A.E. GonÃ§alves.
\newblock Mass and decay spectra of the piecewise uniform string.
\newblock {\em Physics Letters B}, 453(3):217 -- 221, 1999.

\bibitem{lixi91-44-560}
Xz~Li, X~Shi, and Jz~Zhang.
\newblock {Generalized Riemann Zeta-Function Regularization and Casimir Energy
  for a Piecewise Uniform String}.
\newblock {\em Phys.~Rev.~D}, {44}({2}):{560--562}, {1991}.

\bibitem{brev96-53-3224}
Iver~H. Brevik, Holger~Bech Nielsen, and S.D. Odintsov.
\newblock {Casimir energy for a three piece relativistic string}.
\newblock {\em Phys. Rev. D}, 53:3224--3229, 1996.

\bibitem{hada00-62-025011}
L.~Hadasz, G.~Lambiase, and V.~V. Nesterenko.
\newblock Casimir energy of a nonuniform string.
\newblock {\em Phys. Rev. D}, 62:025011, 2000.

\bibitem{brev03-44-1044}
Iver Brevik, Andrei~A. Bytsenko, and Roger Sollie.
\newblock {Thermodynamic properties of the 2N-piece relativistic string}.
\newblock {\em Journal of Mathematical Physics}, 44(3):1044--1055, 2003.

\bibitem{bayi96-37-3662}
Selcuk~Ş. Bayin, J.~P. Krisch, and Mustafa Oezcan.
\newblock The casimir energy of the twisted string loop: Uniform and two
  segment loops.
\newblock {\em Journal of Mathematical Physics}, 37(8):3662--3674, 1996.

\bibitem{bern97-257-84}
M~H Berntsen, I~Brevik, and S~D Odintsov.
\newblock {Casimir theory for the piecewise uniform relativistic string}.
\newblock {\em Ann.~Phys.}, {257}({1}):{84--108}, {1997}.

\bibitem{brev99-40-1127}
I~Brevik, E~Elizalde, R~Sollie, and JB~Aarseth.
\newblock {A new scaling property of the Casimir energy for a piecewise uniform
  string}.
\newblock {\em J.~Math.~Phys.}, {40}({3}):{1127--1135}, {1999}.

\bibitem{schw78-115-1}
J.~Schwinger, L.L. DeRaad, Jr., and K.A. Milton.
\newblock {Casimir Effect in Dielectrics}.
\newblock {\em Ann. Phys.}, 115:1--23, 1978.

\bibitem{bord99-59-085011}
M.~Bordag, K.~Kirsten, and D.V. Vassilevich.
\newblock {On the ground state energy for a penetrable sphere and for a
  dielectric ball}.
\newblock {\em Phys.~Rev.~D}, 59:085011, 1999.

\bibitem{kron31-130-499}
R.~{de}~L. Kronig and W.~G. Penney.
\newblock {Quantum Mechanics of Electrons in Crystal Lattices}.
\newblock {\em Proc.~R.~Soc. A}, {130}:{499}, {1931}.

\bibitem{bord15-91-085038}
M.~Bordag and I.G. Pirozhenko.
\newblock {Surface plasmons for doped graphene}.
\newblock {\em Phys.~Rev.~D}, 91:085038, 2015.

\bibitem{asor06-39-6127}
M.~Asorey, D.~Garcia Alvarez, and J.~M. Munoz-Castaneda.
\newblock {Casimir effect and global theory of boundary conditions}.
\newblock {\em J.~Phys.~A: Math.~Gen.}, {39}:{6127--6136}, {2006}.

\bibitem{bord1812.09022}
Michael Bordag, Jose~M. Mu\~{n}oz Casta\~{n}eda, and Lucia Santamar\'{i}a-Sanz.
\newblock {Vacuum energy for generalised Dirac combs at $T = 0$}.
\newblock arXiv1812.09022, 2018.

\bibitem{bord20-35-2040005}
M.~Bordag.
\newblock {On Bose-Einstein condensation in one-dimensional lattices of delta
  functions}.
\newblock {\em Mod. Phys. Lett.}, A35(03):2040005, 2020.

\bibitem{shaj16-94-065003}
K.~V. Shajesh, Iver Brevik, In\'es Cavero-Pel\'aez, and Prachi Parashar.
\newblock Casimir energies of self-similar plate configurations.
\newblock {\em Phys. Rev. D}, 94:065003, 2016.

\bibitem{BKMM}
M.~Bordag, G.~L. Klimchitskaya, U.~Mohideen, and V.~M. Mostepanenko.
\newblock {\em Advances in the Casimir Effect}.
\newblock Oxford University Press, Oxford, 2009.

\bibitem{geye05-72-022111}
B.~Geyer, G.~L. Klimchitskaya, and V.~M. Mostepanenko.
\newblock {Thermal corrections in the Casimir interaction between a metal and
  dielectric}.
\newblock {\em Phys. Rev. A}, 72:022111, 2005.

\bibitem{milt17-96-085007}
Kimball~A. Milton, Pushpa Kalauni, Prachi Parashar, and Yang Li.
\newblock Casimir self-entropy of a spherical electromagnetic
  $\ensuremath{\delta}$-function shell.
\newblock {\em Phys. Rev. D}, 96:085007, 2017.

\bibitem{bord18-51-455001}
M~Bordag and K~Kirsten.
\newblock On the entropy of a spherical plasma shell.
\newblock {\em J.~Phys.~A: Math.~Gen.}, 51:455001, 2018.

\bibitem{bord1807.10354}
M.~Bordag.
\newblock Entropy in some simple one-dimensional configurations.
\newblock 2018.
\newblock Arxiv: 1807.10354 [quant-ph].

\bibitem{bord20-80-221}
M.~Bordag, J.~M. Munoz-Castaneda, and L.~Santamar\'{i}a-Sanz.
\newblock {Free energy and entropy for finite temperature quantum field theory
  under the influence of periodic backgrounds}.
\newblock {\em Eur.~Phys.~J.~C}, {80}({3}), {MAR 7} {2020}.

\bibitem{brev97-38-2774}
I.~Brevik and R.~Sollie.
\newblock {On the Casimir energy for a 2N-piece relativistic string}.
\newblock {\em Journal of Mathematical Physics}, 38(6):2774--2785, 1997.

\bibitem{grif01-69-137}
D.J. Griffiths and C.A. Steinke.
\newblock {Waves in locally periodic media}.
\newblock {\em {American Journal of Physics}}, {69}({2}):{137--154}, {2001}.

\bibitem{eliz95b}
E.~Elizalde.
\newblock {\em Ten Physical applications of Spectral Zeta Functions}.
\newblock Springer-Verlag, Berlin, 1995.

\bibitem{bord05-38-11027}
M.~Bordag, I.~G. Pirozhenko, and V.~V. Nesterenko.
\newblock Spectral analysis of a flat plasma sheet model.
\newblock {\em J. Phys.}, A38:11027, 2005.

\bibitem{bord08-77-085026}
M.~Bordag and N.~Khusnutdinov.
\newblock On the vacuum energy of a spherical plasma shell.
\newblock {\em Phys.Rev.D}, 77:085026, 2008.

\bibitem{milt19-99-045013}
Kimball~A. Milton, Pushpa Kalauni, Prachi Parashar, and Yang Li.
\newblock {Remarks on the Casimir self-entropy of a spherical electromagnetic
  $\delta$-function shell}.
\newblock {\em Phys. Rev.}, D99(4):045013, 2019.

\bibitem{brev95-51-1869}
I~Brevik and H~B Nielsen.
\newblock {Casimir Theory for the Piecewise Uniform String - Division into 2N
  Pieces}.
\newblock {\em Phys.~Rev.~D}, {51}({4}):{1869--1874}, {1995}.

\bibitem{brev98-15-3383}
I~Brevik, A~A Bytsenko, and H~B Nielsen.
\newblock Thermodynamic properties of the piecewise uniform string.
\newblock {\em Classical and Quantum Gravity}, 15(11):3383--3395, nov 1998.

\end{thebibliography}
\end{document}